\documentclass[fleqn,10pt,showpacs,showkeys]{revtex4}

\usepackage{amsmath}
\usepackage{latexsym}
\usepackage{amssymb}
\usepackage{epsfig,graphicx,psfrag}
\usepackage{afterpage}
\usepackage{mathrsfs}

\begin{document}
%
%
\newcommand{\be}{\begin{equation}} \newcommand{\ee}{\end{equation}}
\newcommand{\ba}{\begin{eqnarray}} \newcommand{\ea}{\end{eqnarray}}
\newcommand{\bea}{\begin{eqnarray}} \newcommand{\eea}{\end{eqnarray}}
\newcommand{\bean}{\begin{eqnarray*}} \newcommand{\eean}{\end{eqnarray*}}
\newcommand{\bom}[1]{\mbox{\boldmath $#1$}}
\newcommand{\s}[1]{{\scriptscriptstyle #1}}
\newcommand{\st}{{\s T}}
\def\slash{\rlap{/}}
%
\title{Non-collinearity in high energy processes\footnote{
Invited talk presented at the Xth Workshop on High Energy Physics
Phenomenology, Chennai (India), January 2-13, 2008}}

\author{P.J.\ Mulders}
\affiliation{VU University - Department of Physics and Astronomy,\\
De Boelelaan 1081, 1081 HV Amsterdam, Netherlands}

\begin{abstract}
We discuss the treatment of intrinsic transverse momenta in high
energy scattering processes. Within the field theoretical framework 
of QCD the process is described in terms of correlators containing
quark and gluon fields. The correlators, parameterized in terms of
distribution and fragmentation functions, contain matrix elements
of nonlocal field configurations requiring a careful treatment to
assure color gauge invariance. It leads to
nontrivial gauge links connecting the parton fields.
For the transverse momentum dependent
correlators the gauge links give rise to time reversal odd phenomena,
showing up as single spin and azimuthal asymmetries. 
The gauge links, arising from multi-gluon initial and final state 
interactions, depend on the color flow in the process, challenging
universality.
\end{abstract}

\maketitle

\section{Introduction}
The basic degrees of freedom that feel the strong interactions, quarks and
gluons, are confined into hadrons, strongly interacting particles.
Considering the nucleons (light hadrons), the characteristic energy and
distance scales are given by the nucleon mass $M_N$, or
taking into account the color degrees of freedom one may prefer a scale
$M_N/N_c \sim$ 300 MeV.
We refer to this as ${\mathscr O}(M)$ or ${\mathscr O}(Q^0)$
if we consider high-energy processes. Such processes are characterised
by hard kinematical variables that are of order $Q$ with 
$Q^2 \gg M_N^2$. Depending on details, the high-energy
scale $Q$ can be the CM energy, $Q \sim \sqrt{s}$ or it can be a
measure of the exchanged momentum.

\begin{figure}[h]
\begin{center}
\epsfig{file=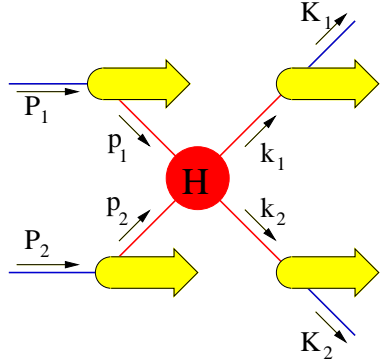,width=5.0cm}
\end{center}
\caption{\label{hardprocess0}
Schematic illustration of the contribution of a hard subprocess,
parton ($p_1$) + parton ($p_2$) $\rightarrow$
parton ($k_1$) + parton ($k_2$),
to the (2-particle inclusive) scattering process
hadron ($P_1$) + hadron ($P_2$) $\rightarrow$
hadron ($K_1$) + hadron ($K_2$) + X,
at the level of the amplitude. The process being hard implies for
the hadronic momenta $P_1\cdot P_2 \sim P_1\cdot K_1 \sim Q^2$, etc.
}
\end{figure}
The basic framework for the strong interactions is Quantum Chromodynamics
(QCD). Hadrons, however,
do not correspond to free particle states created via the quark and
gluon operators in QCD. The situation thus differs from that of QED
with physical electrons and photons. In the latter case
one knows how in the calculation of an S-matrix element contraction
of annihilation and creation operator in the field and particle state
lead to the spinor  wave function. For positive times $\xi^0=t$,
one has
\be
\langle 0\vert \psi_i(\xi)\vert p\rangle =
\langle 0\vert \psi_i(\xi)\,b^\dagger(\bom p)\vert 0\rangle
=\langle 0\vert \psi_i(0)\vert p\rangle \,e^{-i\,p\cdot \xi}
=u_i(\bom p)\,e^{-i\,p\cdot \xi},
\ee
with $p^0 = E_p = \sqrt{\bom p^2+m^2}$.
Such a matrix element is 'untruncated' as seen e.g.\ from
\be
\langle 0\vert \psi_i(\xi)\vert p\rangle\,\theta(t)
= \theta(t)\int \frac{d^4k}{(2\pi)^4}\ e^{-ik\cdot\xi}
\,\frac{i(\slash k + m)}{k^2-m^2+i\epsilon}\,\frac{u_i(\bom p)}{2m}
\,(2\pi)^3\,2E_p\,\delta^3(\bom k-\bom p).
\ee
In a process involving a composite hadronic state $\vert P\rangle$, 
contractions with one or several of the quark and gluon operators may 
be involved, leading to nonzero matrix elements for a quark between the
hadron state and a remainder, but also to nonzero matrix elements
involving multi-parton field combinations,
\[
\langle X\vert \psi_i(\xi)\vert P\rangle,
\ \langle X\vert A^\mu(\eta)\,\psi(\xi)\vert P\rangle, \ldots\ .
\]
For a particular hadron and a parton field combination, one may collect
those operators that involve hadron $\vert P\rangle$ into
(distribution) correlators
\bea
\Phi_{ij}(p;P) & = &
\sum_X\int \frac{d^3P_X}{(2\pi)^3\,2E_X}
\ \langle P\vert \overline\psi_j(0)\vert X\rangle
\,\langle X\vert \psi_i(0)\vert P\rangle\,\delta^4(p+P_X-P)
\nonumber \\ & = &
\frac{1}{(2\pi)^4}\int d^4\xi\ e^{i\,p\cdot \xi}
\ \langle P\vert \overline\psi_j(0)\,\psi_i(\xi)\vert P\rangle ,
\label{qq}
\eea
or correlators involving matrix elements of the form
\be
\Phi^\mu_{ij}(p,p_1;P) =
\frac{1}{(2\pi)^8}\int d^4\xi\,d^4\eta
\ e^{i\,(p-p_1)\cdot \xi}\ e^{i\,p_1\cdot \eta}
\ \langle P\vert \overline\psi_j(0)\,A^\mu(\eta)\,\psi_i(\xi)\vert P\rangle,
\label{qqG}
\ee
pictorially,
\begin{center}
\epsfig{file=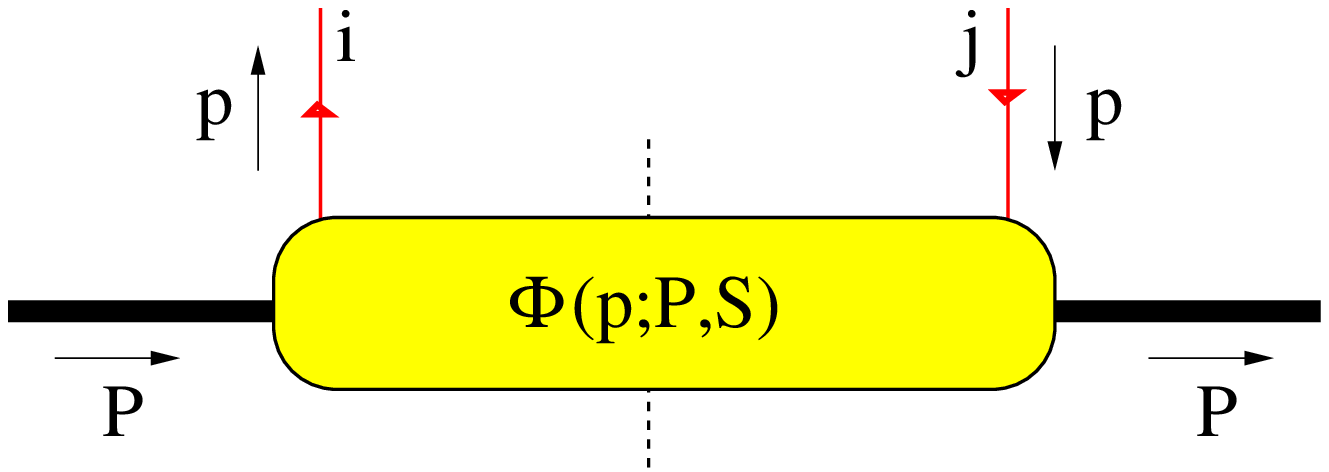,width=5.5cm}
\quad \mbox{or}\quad
\epsfig{file=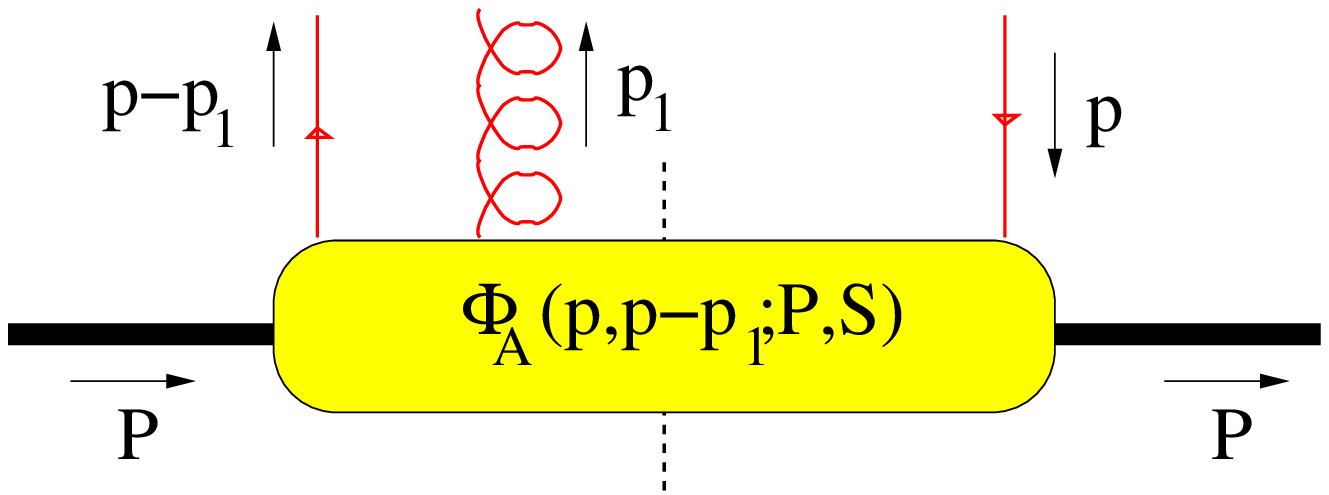,width=5.5cm}.
\end{center}
We will not attempt to calculate these, but leave them as the soft parts,
requiring nonperturbative QCD methods to calculate them. In particular,
although being 'untruncated' in the quark legs, they will no longer
exhibit poles corresponding to free quarks.
These are fully unintegrated parton correlators for initial state hadrons,
in general quite problematic quantities. For example, they are by themselves
not even color gauge-invariant, an issue to be discussed below.
When more hadrons are involved, one could consider two-hadron correlators,
involving two-hadron states (or correlators involving hadronic states
in initial and final state), etc. If the hadrons are well-separated
in momentum phase-space with $P_i\cdot P_j \sim Q^2$, one
expects on dimensional grounds that incoherent contributions
are suppressed by $1/(P_i-P_j)^2 \sim 1/Q^2$ and one can (at least
naively) factorize using forward correlators for single
hadrons, connected by a hard partonic subprocess. 
Such a separation in momentum space requires a hard inclusive scattering
process ($Q^2 \sim s$). The inclusive character is needed to assure that
partons originate from {\em one} hadron, leaving a (target) jet.
In turn, final state partons decay into a jet, in which we also consider
a single identified hadron, which can straightforwardly be extended to a
multi-particle, e.g.\ two-pion, state.
For the fragmentation process of a parton (with momentum $k$) into hadrons
(with momentum $P_h$) we combine the decay matrix elements
in the (fragmentation) correlator, for quarks
\begin{eqnarray}
\Delta_{ij}(k,P_h) & = & \sum_X \frac{1}{(2\pi)^4}
\int d^4\xi\ e^{ik\cdot \xi}\,
\langle 0 \vert \psi_i(\xi) \vert P_h, X \rangle
\langle P_h,X \vert \overline \psi_j(0)
\vert 0 \rangle \nonumber \\
& = & \frac{1}{(2\pi)^4}\int d^4\xi\ e^{ik\cdot \xi}\,
\langle 0 \vert \psi_i (\xi) a_h^\dagger
a_h \overline \psi_j(0) \vert 0 \rangle,
\end{eqnarray}
where an averaging over color indices is implicit.
Pictorially we have
\begin{center}
\epsfig{file=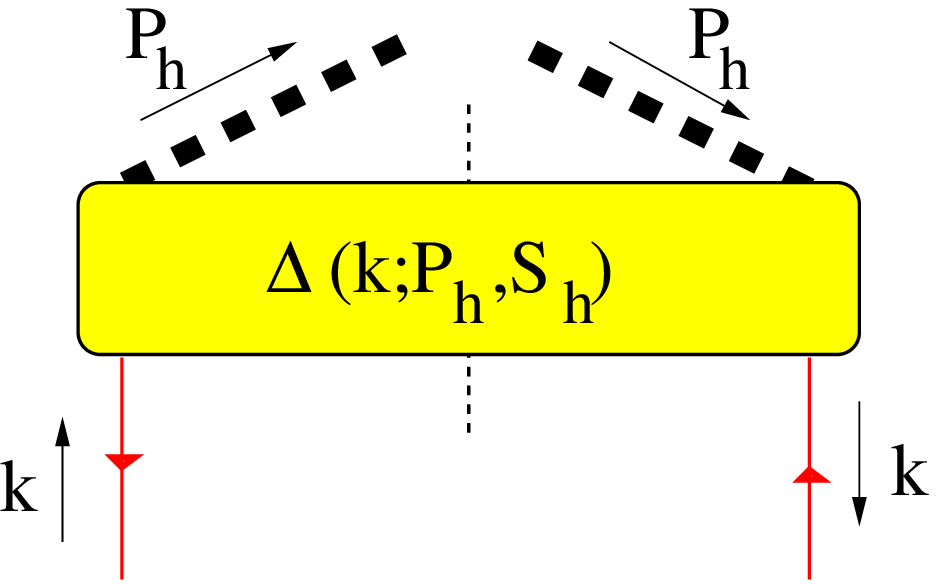,width=4.0cm}
\end{center}
In particular, we note that in fragmentation correlators one no longer
deals with plane wave hadronic states, but with out-states 
$\vert P_h,X\rangle$. In all of the hadronic states mentioned before
one can also consider polarized hadronic states. The spin of quarks is
contained in Dirac structure and that of gluons in the Lorentz 
structure of correlators.

The basic idea in the diagrammatic approach is to realize that the 
correlator involves both hadronic states and quark and gluon operators. 
The correlators can be studied independent from the hard process,
provided we have dealt with the issue of color gauge invariance.
The correlator is the Fourier transform in the space-time
arguments of the quark and gluon fields. In the correlators, all momenta
of hadrons {\em and} quarks and gluons (partons) inside the hadrons are
soft which means that $p^2 \sim p\cdot P \sim P^2 = M_N^2 \ll Q^2 \sim s$.
The {\em off-shellness} being of hadronic order implies that in the
hard process partons are in essence {\em on-shell}.
Consistency of this may be checked by using QCD interactions
to give partons a large off-shellness of ${\mathscr O}(Q)$ and check the
behavior as a function of the momenta. In these considerations one
must also realize that beyond tree-level one has to distinguish bare
and renormalized fields.

\section{\label{TMD}
Collinear and Transverse Momentum Dependent 
Correlators}

In a hard process, the parton fields that appear in the different correlators
correspond to partons in the subprocess for which the momenta satisfy
$p_i\cdot p_j \sim Q^2$. In the study of a particular
correlator it implies the presence of a 'hard' environment.
To connect the correlator to the hard part of the process, it is useful to
introduce for each correlator with hadron momentum $P$,
a null-vector $n$, such that $P\cdot n \sim Q$. Using this relation,
$n$ would be dimensionless.
It is actually more convenient to replace $n/(P\cdot n)$ by a
dimensionful null-vector $n \sim 1/Q$, such that $P\cdot n = 1$.
The vectors $P$ and $n$ can be used to keep track of the importance of
various terms in the correlators and in the components of momentum and
spin vectors.
If one prefers a dimensionless vector, one must make a choice
$P\cdot n \sim Q$. In that case all further appearances of $n$ in
this section should simply be replaced by $n/(P\cdot n)$.
The $n$-vector will acquire
a meaning in the explicit applications or play an intermediary role.
At leading order, it will turn out that the precise form of $n$ doesn't
matter, but at subleading ($1/Q$) order one needs to be careful.

For parton momenta we write the Sudakov decomposition
\be
p = x\,P + p_\st + \underbrace{(p\cdot P - x\,M^2)}_{\sigma}\,n,
\ee
where the term $x\,P \sim Q$, $p_\st \sim M$ and $\sigma\,n \sim M^2/Q$.
We have the exact relations $p\cdot p_{\st}$ = $p_\st^2$ = $(p-x\,P)^2$.
The momentum fraction
$x = p\cdot n$
is ${\mathscr O}(1)$. 

In a hard process, the different importance of the various components
allows up to specific orders in $1/Q$, an integrations over some
components of the parton momenta. The fact that the main contribution
in $\Phi(p;P)$ is assumed to come from regions where $p\cdot P \le M^2$,
whereas the momenta have characteristic scale $Q$,
allows performing the $\sigma$-integration up to $M^2/Q^2$
contributions (and possible contributions from non-integrable tails).
The resulting transverse momentum dependent (TMD) correlators
are light-front correlators,
\be
\Phi_{ij}(x,p_\st;n)  =  \int d(p\cdot P)\ \Phi_{ij}(p;P)
=  \left. \int \frac{d(\xi\cdot P)\,d^2\xi_\st}{(2\pi)^3}
\ e^{i\,p\cdot \xi}
\ \langle P\vert \overline\psi_j(0)\,\psi_i(\xi)\vert P\rangle \right|_{LF}\ ,
\ee
where we have suppressed the dependence on hadron momentum $P$. 
The subscript LF refers to light-front, implying $\xi\cdot n$ = 0.
The light-cone correlators are the correlators containing the
parton distribution functions depending only on the light-cone momentum
fraction $x$,
\be
\Phi_{ij}(x;n)  =  \int d(p\cdot P)\,d^2p_\st\ \Phi_{ij}(p;P)
=\left. \int \frac{d(\xi\cdot P)}{(2\pi)}\ e^{i\,p\cdot \xi}
\ \langle P\vert \overline\psi_j(0)\,\psi_i(\xi)\vert P\rangle \right|_{LC}\ ,
\ee
where the subscript LC refers to light-cone, implying $\xi\cdot n$ = $\xi_\st$
= 0. This integration is generally allowed (again up to $M^2/Q^2$
contributions and contributions coming from tails, e.g.\ logarithmic
corrections from $1/p_\st^2$ tails) if we are interested
in hard processes, in which only hard scales (large invariants
$\sim Q^2$ or ratios thereof, angles, rapidities) are measured.
If one considers hadronic scale observables (correlations or transverse
momenta in jets, slightly off-collinear configurations) one will need
the TMD correlators.

The correlators encompass the information on the soft parts. They
depend on the hadron and (contained) quark momenta $P$ and $p$ (and
spin vectors). The structure of the correlator is reproduced from
these momenta incorporating the required Dirac and Lorentz structure.
Clearly, it is advantageous to maximize the number of components along
the momentum (collinear).  For the {\em soft} scalar objects this means
maximizing the number of contractions with $n$. This leads for nonlocal
operators to the dominance of the twist-2 operators
\be
\overline\psi(0)\slash n\psi(\xi)
\qquad \mbox{and} \qquad
F^{n\alpha}F^{n\beta}(\xi)
\ee
(the latter with transverse indices $\alpha$ and $\beta$).
Twist in this case is just equal to the canonical dimension of the
operator combination (remember that dim($n$) = -1). 

\begin{figure}
\centering
\begin{minipage}{5.5cm}
\centering
\includegraphics[width=4.5cm]{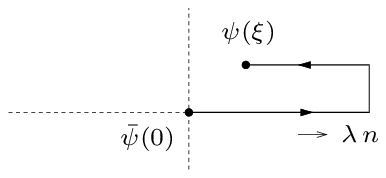}\\
(a) $\Phi^{[+]}{\ \propto\ }
\langle\bar\psi(0)\,U^{[+]}\psi(\xi)\rangle$
\end{minipage}
\hspace{1.0cm}
\begin{minipage}{5.5cm}
\centering
\includegraphics[width=4.5cm]{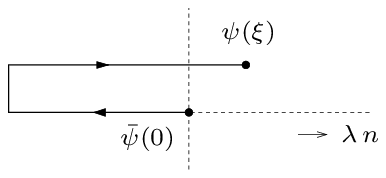}\\
(b) $\Phi^{[-]}{\ \propto\ }
\langle\bar\psi(0)\,U^{[-]}\psi(\xi)\rangle$
\end{minipage}\\[4mm]
\begin{minipage}{5.5cm}
\centering
\includegraphics[width=4.5cm]{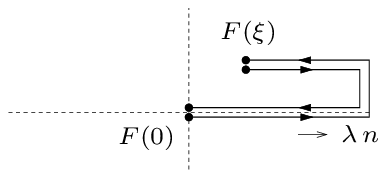}\\
(c) $\Gamma^{[+,+]}{\ \propto\ }
{\rm Tr}\,\langle 
F(0)\,U^{[+]}F(\xi)\,U^{[+]\dagger}\rangle$
\end{minipage}
\hspace{1.0cm}
\begin{minipage}{5.5cm}
\centering
\includegraphics[width=4.5cm]{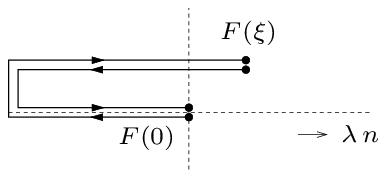}\\
(d) $\Gamma^{[-,-]}{\ \propto\ }
{\rm Tr}\,\langle 
F(0)\,U^{[-]}F(\xi)\,U^{[-]\dagger}\rangle$
\end{minipage}\\[4mm]
\begin{minipage}{5.5cm}
\centering
\includegraphics[width=4.5cm]{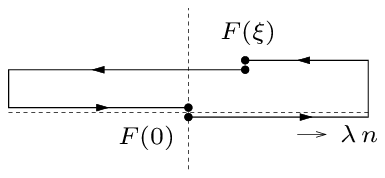}\\
(e) $\Gamma^{[+,-]}{\ \propto\ }
{\rm Tr}\,\langle 
F(0)\,U^{[+]}F(\xi)\,U^{[-]\dagger}\rangle$
\end{minipage}
\hspace{1.0cm}
\begin{minipage}{5.5cm}
\centering
\includegraphics[width=4.5cm]{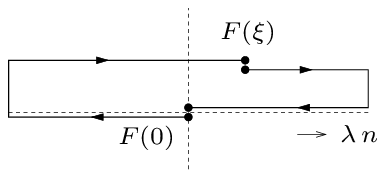}\\
(f) $\Gamma^{[-,+]}{\ \propto\ }
{\rm Tr}\,\langle 
F(0)\,U^{[-]}F(\xi)\,U^{[+]\dagger}\rangle$
\end{minipage}\\[4mm]
\parbox{0.95\textwidth}{\caption{
Simplest structures (without loops)
for gauge links and operators in quark correlators (a)-(b)
and gluon correlators (c)-(f).
\label{simplelinks}}}
\end{figure}

Of course the appearance of the field strength tensor rather than the gauge
field is a requirement of gauge invariance. Besides the field tensor, we
need the inclusion of gauge links 
\be
U^{[n]}_{[0,\xi]}
= {\mathscr P}\,\exp\left(-i\int_0^\xi d(\eta\cdot P)\,n\cdot A(\eta)\right),
\ee
connecting colored parton fields.
In the case of the collinear correlators, the gauge links can be built
from the ${\mathscr O}(1)$ gauge fields $A^+ = A^n = n\cdot A$, 
giving a link along the light-cone ($\xi^+ = n\cdot\xi = \xi_\st = 0$).
The color gauge-invariant light-cone correlators for quarks and gluons are
\bea
&&
\Phi_{ij}(x;n)
= \left. \int \frac{d(\xi\cdot P)}{(2\pi)}
\ e^{i\,p\cdot \xi} \ \langle P\vert
\overline\psi_j(0)\,U^{[n]}_{[0,\xi]}\,\psi_i(\xi)\vert P\rangle \right|_{LC},
\\ &&
\Gamma^{\alpha\beta}(x;n)
= \left. \int \frac{d(\xi\cdot P)}{(2\pi)}
\ e^{i\,p\cdot \xi} \ \langle P\vert
\mbox{Tr}\left(F^{n\beta}(0)\,U^{[n]}_{[0,\xi]}\,F^{n\alpha}(\xi)
\,U^{[n]}_{[\xi,0]}\right)\vert P\rangle \right|_{LC}.
\eea
Using the Taylor expansion of the color gauge-invariant nonlocal operators,
\[
\psi^\dagger(0)\,U_{[0,\xi]}\psi(\xi) = \sum_{n=0}^\infty
(-i)^n\,\frac{x_{\mu_1}\ldots x_{\mu_n}}{n!}
\,\psi^\dagger(0)\,iD^{\mu_1}\ldots iD^{\mu_n}\,\psi(0)
\]
one recovers the irreducible set of symmetric traceless
local operators relevant in the Operator Product Expansion
(OPE) approach to describe $\Phi_{ij}(x)$ and $\Gamma_{\alpha\beta}(x)$, namely
\bean
O_{{\rm quarks}\,ij}^{\mu_1\ldots \mu_n} & = & 
\overline\psi_j(0)\,\gamma^{\{\mu_1}\,iD^{\mu_2}\ldots iD^{\mu_n\}}\,\psi_i(0) 
- \mbox{traces},
\\
O_{{\rm gluons}\,\alpha\beta}^{\mu_1\ldots \mu_n} & = & 
-F_{\beta}^{\ \{\mu_1}(0)\,iD^{\mu_2}\ldots iD^{\mu_{n-1}}
\,F^{\mu_n\}}_{\quad\alpha}(0) - \mbox{traces},
\eean
in which the spin $n$ represents the number of symmetrized indices. 
Subtracting traces is needed to have an irreducible set.
The TMD light-front correlators
\bea
&&
\Phi^{[C]}_{ij}(x,p_\st;n)
= \left. \int \frac{d(\xi\cdot P)\,d^2\xi_\st}{(2\pi)^3}
\ e^{i\,p\cdot \xi} \ \langle P\vert
\overline\psi_j(0)\,U^{[n,C]}_{[0,\xi]}\,\psi_i(\xi)\vert P\rangle \right|_{LF},
\\ &&
\Gamma^{[C,C^\prime]}_{\alpha\beta}(x,p_\st;n)
=  \int \frac{d(\xi\cdot P)\,d^2\xi_\st}{(2\pi)^3}
\ e^{i\,p\cdot \xi}\ \langle P\vert
\mbox{Tr}\left(F^{n}_{\ \ \beta}(0)\,U^{[n,C]}_{[0,\xi]}
\,F^{n}_{\ \ \alpha}(\xi)\,U^{[n,C^\prime]}_{[\xi,0]}\right)
\vert P\rangle \biggr|_{LF},
\eea
involve a more complex link structure, leading to a path dependence
in the definitions (indicated by the arguments $C$ and $C^\prime$).
This arises because of the
(necessary) transverse piece(s) in the gauge link.
The simplest possibilities for the links in the case of quark and
gluon correlators are shown in 
Fig.~\ref{simplelinks}
~\cite{Bomhof:2007xt}.

\section{The Observables}

An important aspect of incorporating intrinsic transverse momenta
is the possibility to access them in experiments. 
Consider the subprocess $\gamma^\ast (q) + q (p) \rightarrow q(k)$ 
which dominates at leading order the inclusive deep inelastic scattering 
(DIS) process $\gamma^\ast (q) + N (P) \rightarrow X$.
In collinear approximation ($p \approx xP$) one finds for the 
momentum fraction $x$ the well-known relation 
$x = x_B = Q^2/2P\cdot q$, i.e.\ the fraction is identified with 
the Bjorken scaling variable. 
For the semi-inclusive deep inelastic scattering (SIDIS) process 
$\gamma^\ast (q) + N (P) \rightarrow h (P_h) + X$ with the same
subprocess one has in collinear approximation ($k \approx P_h/z$) 
the relation $z = z_h = P\cdot P_h/P\cdot q$.
The non-collinearity in this process is determined by
\be
q_\st = q + x_B\,P - \frac{1}{z_h}\,P_h,
\ee 
which is taken zero in collinear approximation. 
First of all, we note that $q_\st$ is experimentally measurable 
being a difference of vectors of ${\mathscr O}(Q)$. This difference
is meaningful at $\mathscr O(M)$ or higher because mass corrections, 
coming from the identification of $x$ with $x_B$ and $z$ with $z_h$, 
appear only at ${\mathscr O}(1/Q^2)$. The vector $q_\st$ is the 
transverse momentum of $q$ in a frame in which $P$ and $P_h$ are chosen 
parallel or (experimentally more useful) it is the transverse momentum 
$-P_{h\perp}/z_h$ in a frame in which $q$ and $P$ are chosen parallel.
We write $Q_\st^2 = -q_\st^2$.
When $Q_\st \sim {\mathscr O}(M)$ one easily sees that the 
Sudakov expansions for the quark momenta, $p\approx xP + p_\st$
and $k\approx \tfrac{1}{z}\,P_h + k_\st$, imply that
$q_\st = k_\st - p_\st$ and the $q_\st$-dependence is attributed
to the (convoluted effect of the) intrinsic transverse momenta in the 
fragmentation and distribution correlators. When $Q_\st \sim
{\mathscr O}(Q)$ a collinear description involving a subprocess with
one additional parton radiated off is needed, but for consistency one
also wants a match with the TMD description~\cite{Bacchetta:2008xw}.

Not only in electroweak processes like SIDIS or the Drell-Yan process
transverse momenta can be accessed. This can also be achieved 
for hadron-hadron scattering. Also here the identification of
the transverse momentum is only possible together with the identification
of the hard subprocess (like pictured in Fig.~\ref{hardprocess0}).
We define
\be
q_\st = \tfrac{1}{z_1}\,K_1 + \tfrac{1}{z_2}\,K_2 
-x_1\,P_1 -x_2\,P_2 = 
p_{1\st} + p_{2\st} - k_{1\st} - k_{2\st},
\ee
a relation valid up to ${\mathscr O}(M)$. The momenta involved
to find $q_\st$ are in principle all ${\mathscr O}(Q)$ and using
at leading order $q_\st \approx 0$ yields relations for the momentum 
fractions in terms of the external hadron momenta 
(up to $1/Q^2$ corrections). The determination
of $q_\st$ at ${\mathscr O}(M)$ gives access to the transverse momenta.
Experimentally one component of $q_\st$ is found as the non-collinearity
of the produced particles $K_1$ and $K_2$ in the plane perpendicular
to the colliding particles $P_1$ and $P_2$, outlined in detail in 
Ref.~\cite{Boer:2003tx}.

Accessing intrinsic transverse momenta in most cases requires a
careful study of azimuthal dependence in high energy processes. Although
the effects are in principle not suppressed by powers of the hard 
scale in comparison with the leading collinear treatment, it 
requires measuring hadronic scale quantities (transverse momenta)
in a high momentum environment. 
Symmetries, in particular time reversal (T) invariance play
an important role: 
\begin{itemize}
\item
The theory of QCD is T-invariant. This makes it sensible to distinguish
quantities and observables according to their T-behavior. 
\item
For distribution correlators involving plane wave hadronic states in the 
definition, combination of the T-operation and hermiticity, shows that 
the collinear correlators $\Phi(x)$ and $\Gamma(x)$ must be T-even. 
For the TMD correlators, however, the T-operation interchanges
$\Phi^{[+]}(x,p_\st) \leftrightarrow \Phi^{[-]}(x,p_\st)$ (and similar
relations for gluon TMD correlators).  
This allows to construct T-even and T-odd combinations.
\item
For fragmentation functions the appearance of an hadronic out-state in
the definition, prohibits
the use of T-symmetry as a constraint and one has always
both T-even and T-odd parts in the correlator (one can refer
to T-even or T-odd in as far as the operator structure is concerned, 
referred to as naive T-even or naive T-odd).
\item
In a scattering process, in which T-symmetry can be used as a constraint,
single spin asymmetries would be forbidden. In fact the only real 
example of this is DIS (omitting electromagnetic interaction
effects). For hadron-hadron scattering, e.g.\ the Drell-Yan process, one 
has a two-hadron initial state and only the assumption of a factorized
description would imply absence of single spin asymmetries.
We now know that this assumption is not valid, even not at leading 
order!
Similarly for processes with identified hadrons in the final state 
T-invariance does not give constraints.
\item
At leading order in $\alpha_s$, however, it is possible to connect
single spin asymmetries (T-odd observable) to the 
T-odd soft parts, since the hard process
will be T-even at this leading order. Collins and Sivers effects as
explanation for single spin asymmetries are the best known examples.
\end{itemize}

\section{The TMD Master Formula}

The description of a hard process is obtained by writing down
hard processes involving quarks and gluons and connecting these to
the soft parts corresponding to initial state hadrons and observed
hadrons in the final state. In the region where the hadrons are 
separated far enough in phase space ($P_i\cdot P_j \sim Q^2$, as 
discussed in the Introduction) one can have a soft part for each of
the hadrons. For the determination of the relevant multi-parton 
matrix elements that need to be included in the calculation
one can use the twist analysis alluded to in Section~\ref{TMD} 
At leading order one needs 
the {\em leading twist} matrix elements 
$\langle \overline\psi \psi\rangle$ and $\langle A_\st A_\st\rangle$, but
also the multi-parton matrix elements 
$\langle \overline\psi \,A^+\ldots A^+\psi\rangle$ and
$\langle A_\st \,A^+\ldots A^+\,A_\st\rangle$ (all having the same twist).
The various matrix elements are resummed into color-gauge invariants 
combinations. Inclusion of the transverse pieces at infinity requires a 
careful analysis~\cite{Belitsky:2002sm}. 
The links that arise are process-dependent. They arise from 
diagrammatic contributions where collinear gluons $A\cdot n$ 
belonging to a particular soft part are attached to parton lines
belonging to a different soft part (which are precisely the external 
parton lines of the linking hard subprocess). 
The link structure, thus, is not affected by inclusion of 
QCD corrections. On the other hand, 
the link structure depends on the color-flow in the specific diagram.

The resulting expression for a hard cross section at measured $q_\st$ is
\be
\frac{d\sigma}{d^2q_\st}\sim \sum_{D,abc\ldots}
\Phi_a^{[C_1(D)]}(x_1,p_{1\st})\,\Phi_b^{[C_2(D)]}(x_2,p_{2\st})
\,\hat\sigma_{ab\rightarrow c\ldots}^{[D]}
\Delta_c^{C_1^\prime(D)]}(z_1,k_{1\st})\ldots + \ldots
\label{basic}
\ee
where the sum $D$ runs over diagrams distinguishing also the
particular color flow and $abc\ldots$ is the summation over quark
and antiquark flavors and gluons. All Dirac and Lorentz indices,
traces etc. are suppressed. The ellipsis at the end indicate 
contributions of other hard processes.

We illustrate this master formula in the following example, taken
from Ref.~\cite{Bomhof:2006dp}, describing the contribution in a
hard scattering process coming from the $qq\rightarrow qq$ subprocess
(with both quarks having the same flavor).
There are four diagrammatic contributions (see Table~\ref{Tqq2qq}), 
the result of which can be denoted as $\hat \sigma_{qq\rightarrow qq}^{[D]}$
with $D$ running over the diagrams. For the first diagram,
there are two different possibilities for the color flow, which 
absorbing the overall color factor in $\hat \sigma$ have strengths
$(N_c^2+1)/(N_c^2-1)$ and $-2/(N_c^2-1)$, respectively. The second diagram
has the same color flow possibilities. The third and fourth diagrams also
have identical color flow possibilities but different from the first
two diagrams.
\begin{table}[t]
\begin{center}
\includegraphics{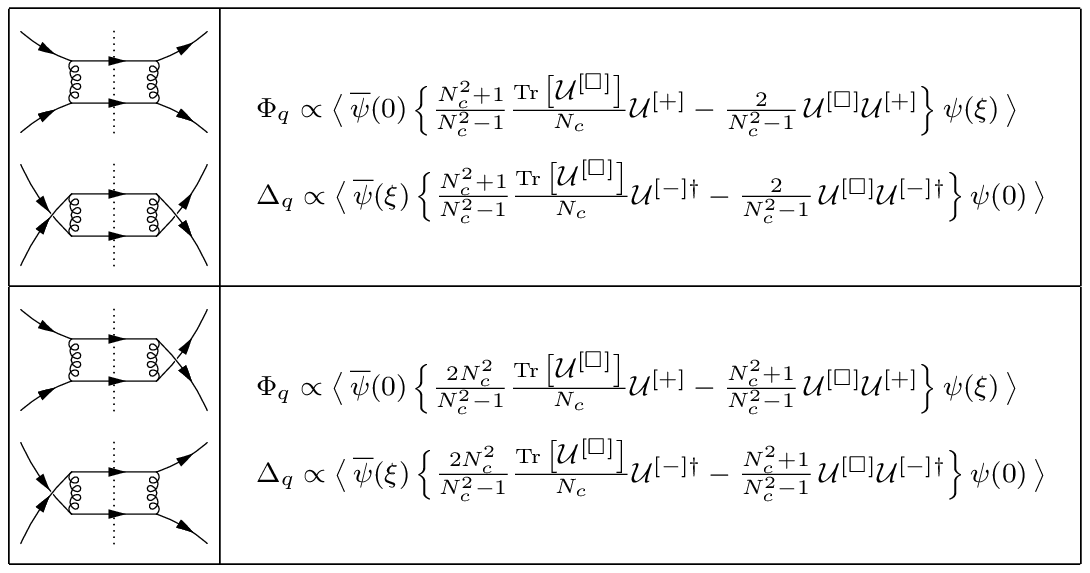}
\end{center}
\caption{
Gauge-links appearing in the soft parts connected to the 
$qq\rightarrow qq$ subprocess depend on the specific 
diagrams.
The paths for $\Phi_q(x,p_\st)$ are shown in Fig.~\ref{qqlinks}.
\label{Tqq2qq}
}
\end{table}
\begin{figure}
\begin{center}
\epsfig{file=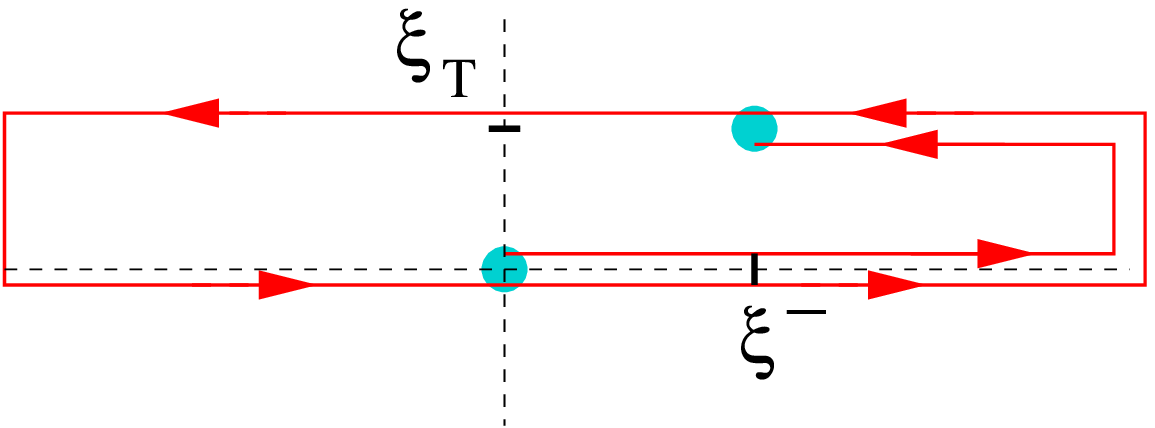,width=4.5cm}
\hspace{1.5cm}\epsfig{file=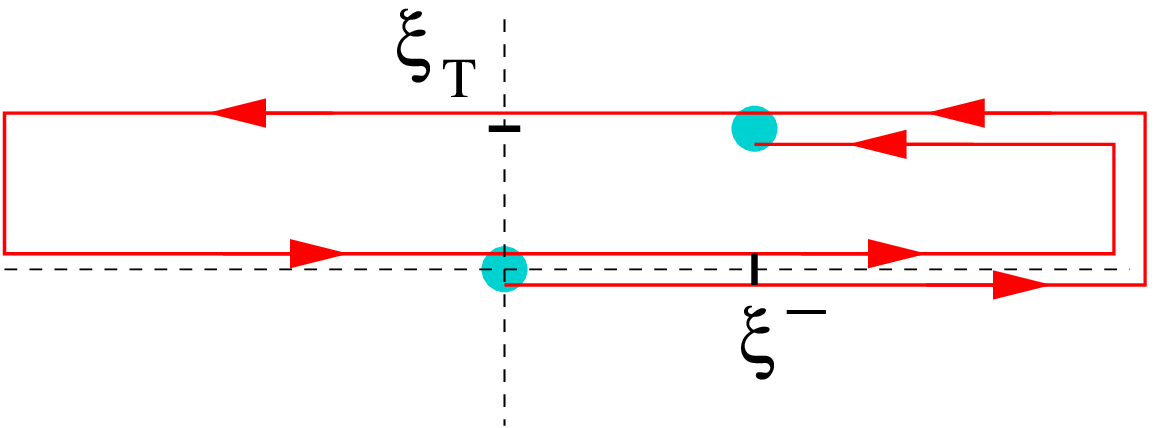,width=4.5cm}
\end{center}
\caption{\label{qqlinks}
The paths in the gauge links in $\Phi_q(x,p_\st)$. They involve a 
loop $U^{[\Box]}$ = $U^{[+]}\,U^{[-]\dagger}$,
which in the path shown in the left figure is closed (color-trace) and in
the right figure is followed by a $[+]$-path. We will use
the shorthand notations $\Phi^{[(\Box)+]}$ and $\Phi^{[\Box +]}$ respectively
with $(\Box)$ indicating the color-tracing and averaging.
}
\end{figure}
In this case each of the diagrams contributes two terms to the sum in
Eq.~\ref{basic}, e.g.\ the first diagram yields
\bea
&&\frac{d\sigma}{d^2q_\st}\sim 
\Phi_q^{[(\Box)+]}(1)\,\Phi_q^{[(\Box)+]}(2)
\,\underbrace{\frac{N_c^2+1}{N_c^2-1}\,\hat\sigma_{qq\rightarrow qq}^{[D_1]}}
\Delta_q^{[(\Box)-^\dagger]}(1^\prime)
\,\Delta_q^{[(\Box)-^\dagger]}(2^\prime)
\nonumber\\
&&\mbox{}\hspace{1cm}
+\Phi_q^{[\Box +]}(1)\,\Phi_q^{[\Box +]}(2)
\,\underbrace{\frac{-2}{N_c^2-1}\,\hat\sigma_{qq\rightarrow qq}^{[D_1]}}
\Delta_q^{[\Box -^\dagger]}(1^\prime)
\,\Delta_q^{[\Box -^\dagger]}(2^\prime) + \ldots,
\eea
where the underbraced items are separate $\hat\sigma^{[D]}$-entries in
the $D$-summation of Eq.~\ref{basic}.

\section{Integrated and weighted cross section}

The results for cross sections after integration over the transverse
momenta $q_\st$ involve the path-independent integrated correlators 
$\Phi(x)$ rather than
the path-dependent TMD correlators $\Phi^{[C(D)]}(x,p_\st)$. Thus, from
Eq.~\ref{basic} one gets the well-known result
\be
\sigma \sim \sum_{abc\ldots} \Phi_a(x_1)\,\Phi_b(x_2)
\,\hat\sigma_{ab\rightarrow c\ldots}\Delta_c(z_1)\ldots + \ldots,
\ee
where
\be
\hat \sigma_{ab\rightarrow c\ldots} 
= \sum_{D} \hat\sigma^{[D]}_{ab\rightarrow c\ldots}
\label{parton}
\ee
is the partonic cross section.

Constructing a weighted cross section (azimuthal asymmetry) by 
including a weight $q_\st^\alpha$ in the $q_\st$-integration leads,
with the help of the relation between the observable $q_\st$ and the intrinsic
transverse momenta (e.g.\ the relation $q_\st = p_\st-k_\st$ in 
SIDIS), to soft correlators of the form
\be
\Phi_\partial^{\alpha\,[C]}(x) =
\int d^2p_\st\ p_\st^\alpha\Phi^{[C]}(x,p_\st).
\label{moment}
\ee
These still contain a path dependence, so Eq.~\ref{basic} cannot be
simplified immediately. However, it turns out that the correlator
in Eq.~\ref{moment} can be expressed as
\be
\Phi_\partial^{\alpha\,[C]}(x) =
\widetilde\Phi_\partial^{\alpha}(x) 
+ C_G^{[U(C)]}\,\pi\Phi_G^\alpha(x,x).
\label{decomposition}
\ee
Here $\widetilde\Phi_\partial(x)$ is a collinear correlator
containing matrix elements with T-even operators, while $\Phi_G(x,x_1)$ is
a collinear correlator with a structure like the quark-gluon-quark 
correlator shown in Eq.~\ref{qqG} involving the gluon field $F^{n\alpha}$. 
In Eq.~\ref{decomposition} one needs the zero-momentum ($x_1 = 0$) limit
for the gluon momentum. This matrix element is known as the gluonic pole
matrix element. The operators involved are T-odd. Both collinear
correlators on the RHS in Eq.~\ref{decomposition} are link-independent.
The gluonic pole factors $C_G$ multiplying the gluonic pole correlator in 
Eq.~\ref{decomposition}, however, do depend on the gauge link. They
can be easily calculated. We have
for instance $C_G^{[\pm]} = \pm 1$, $C_G^{[\Box +]} = 3$
and $C_G^{[(\Box)+]} = 1$.
Thus, one can write for the single-weighted cross section
\bea
\left< q_\st^\alpha\sigma\right> & = &
\int d^2q\st\ q_\st^\alpha \frac{d^2\sigma}{d^2q\st} 
= \sum_{D,abc\ldots}\Phi_{\partial\,a}^{\alpha\,[C]}(x_1)\,\Phi_b(x_2)
\,\hat\sigma^{[D]}_{ab\rightarrow c\ldots}\Delta_c(z_1)\ldots + \ldots
\nonumber \\ & = &
\sum_{abc\ldots}\widetilde\Phi_{\partial\,a}^{\alpha}(x_1)\,\Phi_b(x_2)
\hat\sigma_{ab\rightarrow c\ldots}\Delta_c(z_1)\ldots + \ldots
\nonumber \\ &&\mbox{}\hspace{0.5cm}
+ \sum_{abc\ldots}\pi\Phi_{G\,a}^{\alpha}(x_1,x_1)\,\Phi_b(x_2)
\,\hat\sigma_{[a]b\rightarrow c\ldots}\Delta_c(z_1)\ldots + \ldots
\eea
where the first term is multiplied by the normal parton cross section
(Eq.~\ref{parton}) and the second one involves the {\em gluonic
pole cross section},
\be
\hat \sigma_{[a]b\rightarrow c\ldots} 
= \sum_{D} C_G^{[U(C(D))]}\hat\sigma^{[D]}_{ab\rightarrow c\ldots}\ .
\label{gpxs}
\ee
Noteworthy is the fact that these gluonic pole cross sections like
the normal partonic cross sections also
constitute gauge invariant combinations of the squared amplitudes.
While for the electroweak processes like SIDIS and DY one has
a simple factor, 
$\hat\sigma_{\ell [q]\rightarrow \ell q} 
= +\hat\sigma_{\ell q\rightarrow \ell q}$
and
$\hat\sigma_{[q]\bar q\rightarrow \ell \bar\ell} 
= -\hat\sigma_{q\bar q\rightarrow \ell \bar\ell}$,
the result for $qq\rightarrow qq$ is more complex,
\bea
\hat\sigma_{qq\rightarrow qq} & = &
\hat\sigma^{[D_1]} + \hat\sigma^{[D_2]}
+\hat\sigma^{[D_3]} + \hat\sigma^{[D_4]},
\\
\hat\sigma_{[q]q\rightarrow qq} & = &
\frac{N_c^2-5}{N_c^2-1}\left(\hat\sigma^{[D_1]} + \hat\sigma^{[D_2]}\right)
-\frac{N_c^2+3}{N_c^2-1}\left(\hat\sigma^{[D_3]} + \hat\sigma^{[D_4]}\right),
\eea
where $\hat\sigma^{[D_i]}$ refer to the contributions coming from the
diagrams in Table~\ref{Tqq2qq}. Actually the results simplify in the limit
$N_c\rightarrow \infty$ in which case the color flow is unique for each
diagram. Explicit results for gluonic pole cross sections are given 
in Ref.~\cite{Bomhof:2006ra}.

The approach to understand T-odd observables like single spin asymmetries 
via the TMD correlators and the non-trivial gauge link structure unifies
a number of approaches to understand such observables, in particular
the collinear approach of Qiu and Sterman~\cite{Qiu:1991pp} 
and the inclusion of soft gluon interactions 
by Brodsky and collaborators~\cite{Brodsky:2002cx}.
Although the treatment of fragmentation correlators also separates into
parts with T-even and T-odd operator structure, gluonic pole
contributions (T-odd parts) in the case of fragmentation might vanish.
Indications come from the soft-gluon approach~\cite{Collins:2004nx} 
and a recent spectral analysis in a spectator model 
approach~\cite{Gamberg:2008yt}. 

\section{Universality}

Clearly the TMD master formula in Eq.~\ref{basic} breaks universality
in the sense that one needs to know TMD correlators $\Phi^{[U]}(x,p_\st)$
with all sorts of gauge links $U$. Certainly it would be desirable to
perform a study of the effects on the
soft correlators caused by more complex gauge links than the simple ones 
given in Fig.~\ref{simplelinks}.

A useful procedure is to rewrite the TMD correlators in terms of
T-even and T-odd correlators constructed from those
in Fig.~\ref{simplelinks} and a residual or {\em junk} part,
\be
\Phi^{[U]}(x,p_\st) = 
\frac{1}{2}\Phi^{[{\rm even}]} + \frac{1}{2}\,C_G^{[U]}\Phi^{[{\rm odd}]}(x,p_\st)
+ \delta\Phi^{[U]}(x,p_\st),
\ee
which by construction leads to
\bea
&&\Phi^{[{\rm even}]}(x) = \Phi(x),
\quad
\Phi_\partial^{\alpha\,[{\rm even}]}(x) = \widetilde\Phi^\alpha_\partial(x)
\\
\quad
&&\Phi^{[{\rm odd}]}(x) = 0,
\qquad\quad
\Phi_\partial^{\alpha\,[{\rm odd}]}(x) = \pi\,\Phi^\alpha_G(x,x)
\\
&&\delta\Phi^{[U]}(x) = 0,
\qquad\ \quad
\delta\Phi_\partial^{\alpha\,[U]}(x) = 0.
\eea
For the simple quark correlators one has $\delta\Phi^{[+]} =
\delta\Phi^{[-]} = 0$ and the T-even and T-odd combinations are
\bea
&&\Phi^{[{\rm even}]}(x,p_\st) =
\tfrac{1}{2}\left(\Phi^{[+]}(x,p_\st) + \Phi^{[-]}(x,p_\st)\right) 
\\
&&\Phi^{[{\rm odd}]}(x,p_\st) =
\tfrac{1}{2}\left(\Phi^{[+]}(x,p_\st) - \Phi^{[-]}(x,p_\st)\right) .
\eea 
Then one finds for instance 
\bea
\Phi^{[\Box +]}(x,p_\st) & = &
\tfrac{1}{2}\,\Phi^{[{\rm even}]}(x,p_\st) + \tfrac{3}{2}\,\Phi^{[{\rm odd}]}(x,p_\st)
+ \delta\Phi^{[\Box +]}(x,p_\st),
\nonumber
\\ & = &
2\,\Phi^{[+]}(x,p_\st) - \Phi^{[-]}(x,p_\st)
+ \delta\Phi^{[\Box +]}(x,p_\st).
\nonumber
\eea
In Ref.~\cite{Bomhof:2007xt} it was noted that some further 
non-trivial simplification
occurs for quark junk TMD while also the simple gluon correlators 
(Fig.~\ref{simplelinks}) can be regrouped into 
T-even and T-odd combinations,
\bea
&&\Gamma^{[{\rm even}]}(x,p_\st)
= \tfrac{1}{2}\,\Gamma^{[+,+]}(x,p_\st)
+ \tfrac{1}{2}\,\Gamma^{[-,-]}(x,p_\st),
\\
&&\Gamma_F^{[{\rm odd}]}(x,p_\st)
= \tfrac{1}{2}\,\Gamma^{[+,+]}(x,p_\st)
- \tfrac{1}{2}\,\Gamma^{[-,-]}(x,p_\st),
\\
&&\Gamma_D^{[{\rm odd}]}(x,p_\st)
= \tfrac{1}{2}\,\Gamma^{[+,-]}(x,p_\st)
- \tfrac{1}{2}\,\Gamma^{[-,+]}(x,p_\st).
\eea
The two T-odd correlators reduce to the two three-gluon gluonic pole 
correlators $\Gamma^F_G(x,x)$ and $\Gamma^D_G(x,x)$, which differ
in the way the three color octets are coupled to a color singlet.

The contributions $\delta\Phi$ and $\delta\Gamma$ make the non-universality
explicit, which is a first step if one wants to study and possibly prove
factorization in the case of TMD correlators. For phenomenological
studies a reasonable first step is to omit the junk TMD, knowing that 
they will average to zero in a weighted asymmetry. 
This approximation can be applied immediately in the TMD master formula
(Eq.~\ref{basic}) in which the color flow possibilities are distinguished.
This master formula remains the basic starting point, used in recent analyses
of photon-jet~\cite{Boer:2007nd} and jet-jet~\cite{Lu:2007}.
production in hadron-hadron scattering.
In the case of a linear weighting with the transverse momentum one 
can conveniently cast the result into folding of T-even and T-odd 
functions with normal and gluonic pole partonic cross sections, 
respectively. The procedure has been investigated for several 
processes~\cite{Bomhof:2007su},
in particular comparing the effect of using normal versus gluonic
pole cross sections.


\begin{thebibliography}{99}
\bibitem{Bomhof:2007xt}
For details and references, see
C.J. Bomhof and P.J. Mulders,
Nucl. Phys. B795 (2008) 409.
\bibitem{Bacchetta:2008xw} 
A. Bacchetta, D. Boer, M. Diehl and P.J. Mulders, 
e-Print: arXiv:0803.0227 [hep-ph].
\bibitem{Boer:2003tx}
D. Boer and W. Vogelsang, 
Phys. Rev. D69 (2004) 054028;
A. Bacchetta, C.J. Bomhof, P.J. Mulders and F. Pijlman,
Phys. Rev. D72 (2005) 034030. 
\bibitem{Belitsky:2002sm} 
A.V. Belitsky, X. Ji and F. Yuan, 
Nucl. Phys. B656 (2003) 165; 
D. Boer, P.J. Mulders and F. Pijlman,
Nucl. Phys. B667 (2003) 201.
\bibitem{Bomhof:2006dp}
C.J. Bomhof, P.J. Mulders and F. Pijlman,
Eur. Phys. J. C47 (2006) 147.
\bibitem{Qiu:1991pp}
J.-W. Qiu and G. Sterman,
Phys. Rev. Lett. 67 (1991) 2264;
Nucl. Phys. B378 (1992) 52.
\bibitem{Brodsky:2002cx}
S.J. Brodsky, D.S. Hwang and I. Schmidt, 
Phys. Lett. B530 (2002) 99,
Nucl. Phys. B642 (2002) 344.
\bibitem{Bomhof:2006ra}
C.J. Bomhof and P.J. Mulders, JHEP 82 (2007) 029.
\bibitem{Collins:2004nx}
J.C. Collins and A. Metz,
Phys. Rev. Lett. 93 (2004) 252001.
\bibitem{Gamberg:2008yt}
L.P. Gamberg, A. Mukherjee and P.J. Mulders, 
e-Print: ArXiv:0803.2632 [hep-ph],
to be publ. in Phys. Rev. D.
\bibitem{Boer:2007nd}
D. Boer, P.J. Mulders and C. Pisano,
Phys. Lett. B660 (2008) 360.
\bibitem{Lu:2007}
Z. Lu and I. Schmidt,
e-Print: arXiv:0805.4006 [hep-ph].
\bibitem{Bomhof:2007su}
C.J. Bomhof, P.J. Mulders, W. Vogelsang and F. Yuan,
Phys. Rev. D75 (2007) 074019;
A. Bacchetta, C. Bomhof, U. D'Alesio, P.J. Mulders and F. Murgia,
Phys. Rev. Lett. 99 (2007) 212002;
D. Boer, C.J. Bomhof, D.S. Hwang and P.J. Mulders,
Phys. Lett. B659 (2008) 127.
\end{thebibliography}
\end{document}